\documentclass[useAMS,usenatbib,usegraphicx]{mn2e}
\usepackage{colortbl}
\definecolor{red2}{rgb}{1.0,0.0,0.} 
\definecolor{lightblue}{rgb}{0.25,0.,0.95}
\def \hs{\color{red2}\bf}

\def \hs{\color{red2}}

\title[Barlenses]{Milky Way mass galaxies with X-shaped bulges are not
  rare in the local Universe} \author[Laurikainen E. et
  al.]{Laurikainen E.$^{1},^{2}$\thanks{E-mail:
    eija.laurikainen@oulu.fi}, Salo H.$^{1}$, Athanassoula E.$^{3}$,
  Bosma A.$^{3}$, Herrera-Endoqui M.$^{1}$ \\ $^{1}$Dept of
  Physics/Astronomy Division, University of Oulu, FIN-90014, Finland
  \\ $^{2}$Finnish Centre of Astronomy with ESO (FINCA), University of
  Turku, V\"ais\"al\"antie 20, FI-21500 Piikki\"o, Finland \\ $^{3}$ Aix Marseille Universit\'e, CNRS, LAM (Laboratoire d'Astrophysique
de Marseille) UMR 7326, 13388, Marseille, France}

\begin{document}

\date{Accepted:  Received:  }

\pagerange{\pageref{firstpage}--\pageref{lastpage}} \pubyear{2014}

\maketitle

\label{firstpage}

\begin{abstract}

Boxy/Peanut/X-shaped (B/P/X) bulges are studied using the 3.6 $\mu$m
images from the Spitzer Survey of Stellar Structure in Galaxies
(S$^4$G), and the K$_s$-band images from the Near-IR S0 galaxy Survey
(NIRS0S). They are compared with the properties of barlenses, defined
as lens-like structures embedded in bars, with sizes of $\sim$50$\%$
of bars and axial ratios of $\sim$ 0.6-0.9.  Based on observations
(extending Laurikainen et al.) and recent simulation models
(Athanassoula et al.) we show evidence that barlenses are the
more face-on counterparts of B/P/X-shaped bulges. Using unsharp masks
18 new X-shaped structures are identified, covering a large
  range of galaxy inclinations. The
  similar masses and red B-3.6$\mu$m colors of the host galaxies, and
  the fact that the combined axial ratio distribution of the host galaxy
  disks is flat, support the interpretation that barlenses and
  X-shapes are physically the same phenomenon. 
In Hubble types $-3 \leq T \leq 2$ even
  half of the bars contain either a barlens or an X-shaped structure. 
Our detailed 2D
  multi-component decompositions for 29 galaxies, fitting the
  barlens/X-shape with a separate component, indicate very small or
  non-existent classical bulges. Taking into account that the
structures we study have similar host galaxy masses as the Milky
Way (MW), our results imply that MW mass galaxies with no significant
classical bulges are common in the nearby Universe.
\end{abstract}


\section{Introduction}

Understanding the formation of bulges is a key in any model of galaxy
formation and evolution. The main types of bulges are classical bulges
formed in the primordial galaxies or by galaxy mergers, and bulges
formed by secular processes \citep{kormendy2004}.  One type of
non-classical bulges forms via bar induced gas inflow, followed by star
formation, which can be interpreted as small inner disks (``disky
bulges'' by Athanassoula 2005). Or they can be Boxy/Peanut (B/P)
bulges \citep{combes1981,atha2002}, being actually vertically
thick inner parts of bars seen edge-on \citep{luttice2000}.  In
unsharp mask images they often show X-shapes \citep{bureau2006}, and
are collectively called as B/P/X-shaped structures.

The ``disky bulges'' have apparent flattening similar to that of
the disk \citep{kent1985}, blue optical colors, and S\'ersic
indexes indicative of nearly exponential surface brightness profiles
\citep{drory2007}. B/P/X bulges form part of the bar and therefore are
expected to have similar colors as the inner disk.  Classical bulges have
old stellar populations and enhanced [$\alpha$/Fe] abundances
\citep{macarthur2009,sanchez2011}.  However, it is not easy to
distinguish classical and secularly formed bulges based on their
colors or ages. Bulges can be re-juvenated by inflow of gas via cosmic
filaments, wet minor mergers, or cooling gas from the baryonic halos,
giving a mixture of stellar ages and abundances
\citep{coelho2011,perez2009}, and complex stellar kinematics
\citep{mendez2008,williams2011}.  The mechanisms that re-juvenate
bulges can also make them flatter and faster rotating
\citep{Khochfar2011}.
Classical bulges, covering a large range of bulge-to-total ($B/T$) flux ratios,
can be qualitatively reproduced by the current cosmological simulations
within the hierarchical galaxy assembly \citep{kauffman1996,governato2007}.

Nevertheless, more quantitative analysis has shown that in
cosmological simulations the cumulative effects of adding mass to the
bulge and increasing its S\'ersic index during the galaxy evolution
are easily overestimated (Silk, Di Cintio $\&$ Dvorkin 2013 and references therein).  Many high
redshift galaxies, in the environment where bulge formation is
expected to be vigorous, show clumpy morphologies without any clear
central mass concentrations
(F\"orster Schreiber et al. 2009, 2011). 
Also, analysis of the $B/T$ flux-ratios
in the local Universe  (Laurikainen et al. 2005, 2010, Graham $\&$
Worley 2008, Gadotti 2009, Kormendy et al. 2010, Weinzirl et al. 2009)
show systematically less massive bulges than predicted by 
cosmological simulations.  

In this study we take a new approach to analyze bulges/inner bar
components, deriving even stricter limit for the fraction of light
in classical bulges. The central ingredient is to correctly
  account for the {\it barlens} (bl) component, as its light is often
erroneously attributed to a classical bulge. Barlenses, e.g. lens-like
structures embedded in bars were identified as distinct structural
features by Laurikainen et al. (2007, 2011), where it was also
speculated that they are vertically extended features forming the
inner parts of bars. This conjecture is shown to be true
    for the barlens component of simulated galaxies in Athanassoula et
    al. (2014).  In this letter we give further evidence
  showing that observed barlenses, appearing in fairly face-on
  galaxies, are physically the same phenomenon as the B/P/X-shaped
  bulges in more inclined galaxies.

\section{Sample and data analysis}

\subsection{Sample}

 
We use 3.6 $\mu$m images of the Spitzer Survey of Stellar Structure in
Galaxies (S$^4$G, Sheth et al. 2010), which is a sample of 2352
nearby galaxies, covering all Hubble types and disk inclinations with
total blue magnitude B$_T \le $ 15.5 mag. Our other database is the
Near-IR S0-Sa galaxy Survey (NIRS0S, Laurikainen et al. 2011) of 
206 early-type disk galaxies observed at K$_s$-band, with $i$ $\le$
65$^\circ$.  Because S$^4$G contains only
galaxies with HI emission, a sample combined with NIRS0S guarantees
that also early-type disk galaxies lacking gas are included: NIRS0S
contains 113 galaxies not in S$^4$G, leading to the total number of 2465
galaxies.  As we are interested in the relative masses of the
structural components, near and mid-IR images are used, tracing the old stellar
populations (see Meidt et al. 2012).

  Our combined sample contains 80 barlenses and 89 X-shaped (or
   B/P) structures, based on the morphological classifications by \citet{lauri2011}
   and Buta et al. (2014), or found in our unsharp mask analysis
   described below. These numbers not necessarily contain all bulges with boxy
   isophotes.  In our statistical analysis we limit to $B_T \le 12.5$
   mag, which is the NIRSOS completeness limit: this leaves $N=597$
   galaxies (99 from NIRSOS): 365 are barred (SB or
   SAB), 42 host X-shapes and 61 have barlenses, indicating an overall
   X/bl fraction of 28$\%$ among barred galaxies.
As NIRS0S galaxies have $i \le 65^\circ$ we
   probably lack some gas poor high inclination systems with
   B/P/X-shapes.  Based on the frequency of B/P/X-shapes among highly
   inclined early-type S$^4$G galaxies, we estimate that at most
   $\sim$10 such galaxies are missing from our mag-limited
   subsample.

%

\subsection{Unsharp masking and multi-component decompositions}
In order to be sure that all X-shapes are found, unsharp
masks were created for all S$^4$G and NIRS0S galaxies.  We convolved the
images with a Gaussian kernel (width 5-20 pix), and divided
the original image with the convolved image. As a result 18 new
X-shapes were discovered.\footnote{The new X-shape detections are:
  ESO404-027, ESO443-042, IC0335, IC1067, IC3806, NGC0522, NGC0660,
  NGC1476, NGC3098, NGC4856, NGC5022, NGC5145, NGC5375, NGC5806,
  NGC7513, NGC3692, NGC5145, NGC5757. Some of these galaxies were
  identified as B/P in Buta et al. (2014).}

We perform detailed multi-component decompositions on 30 galaxies
(with $i \leq 65^\circ$), of which 15 have barlenses and 14 X-shaped
structures. Separate components are included for the inner (bl/B/P/X)
and the outer thin bar components, and the main emphasis is to see how
this affects the derived bulge component. The decompositions were
carried out using GALFIT \citep{peng2010}, with the help of
IDL-based GALFIDL-procedures \citep{salo2014}.  We use a S\'ersic
function for the bulge, disk, and the inner bar components, and
Ferrers function for the outer bar. If needed, an extra unresolved
central source is also added.  In distinction to the starting models,
taken from Laurikainen et al. (2010) for NIRS0S, and from Salo et
al. (2014) for S$^4$G, the shape parameters $\alpha$ and $\beta$ in
the Ferrers function, and the boxy/disciness of the isophotes were
all left free in the fit.  The use of a S\'ersic function for the disk
guaranteed that also Freeman type II profiles with disk breaks or
non-exponential surface brightness profiles are fitted in a
reasonable manner.

\section{Results and discussion}

\subsection{Barlenses - face-on counterparts of B/P/X-shaped ``bulges''}

B/P/X-shaped bulges \citep{combes1981,pfenn1991,atha2005} have been
extensively discussed in galaxies seen edge-on.  Simulation models
have also shown that B/P-bulges can develop X-shapes when the bar
grows in strength (see Athanassoula 2005).  Athanassoula $\&$ Beaton
(2006) showed that such bulges are visible not only in the edge-on
geometry, but also in fairly inclined galaxies, using isophotal
analysis. This was extended \citet{erwin2013} to
moderately inclined galaxies, applied to a sample of 78 galaxies.

Barlenses (though not yet called as such) were speculated to be
  the nearly face-on counterparts of B/P/X-shaped structures by
\citet{lauri2007}. In distinction to classical bulges
  barlenses have flatter light distributions, and in distinction
  to nuclear lenses they are larger, covering $\sim$50$\%$ of the bar
  size (Laurikainen et al. 2011, 2013).
No theoretical proof
  for this existed at that time.  In Athanassoula et
  al. (2014) we identify barlens components in N-body + SPH
  simulations, and find a good agreement with observed
  barlenses. Moreover, by comparing views from different
  directions, this study shows that the simulated barlenses and
  B/P/X bulges are indeed the same component.

 An example of the good correspondence between
 simulations/observations is illustrated in Figure 1, which compares
 the NIRS0S image of NGC 4314 (a prototypical barlens galaxy) with a
 simulation model from Athanassoula et al. (2013).  Along the bar major axis
 the barlens connects smoothly into the thin bar.  There is also a
 central peak which in principle could be a separate small bulge.
 However, at least in this case it is clear that the bulge is by no
 means a classical bulge, as the 'bulge' region contains several
 disk-specific structures (see Benedict et al. 1993, 2002; Erwin $\&$
 Sparke 2003). A comparison with the simulation model also shows a
 remarkable similarity of the profiles, including the central peak
 which in the model forms part of the bar (the model does not include any
 separate bulge component).  

As a further evidence supporting this conjecture, we find that
the combined axial ratio distribution of the galaxies with barlenses
and X-shaped structures is flat (Fig. 2a), suggesting that these
structures indeed are physically the same phenomenon. Consistent
with this idea is that both types of bars reside in massive galaxies
(Fig. 2b) with similar red $B$-3.6 $\mu$m (AB) colors. The median
masses of the  host galaxies are log M*/M$_{\odot}$=10.57 and 10.52,
and the colors 1.25 and 1.26, for the barlens and X-shaped galaxies,
respectively.
Barlenses appear shifted towards earlier Hubble types (see Fig. 2c; MW
  has T=3), but most likely this is simply 
 a matter of classification bias: barlenses are easily
 confused with classical bulges and thus the hosts are assigned an earlier type.  This
  is in accordance with the simulation models (Athanassoula et al. 2014, see Fig. 1 lower left panel) which
  suggest that a given model may appear to possess an extended round
  central component in near face-on view, which however is much less
 pronounced when viewed edge-on.
Interestingly, the masses in Figure 2b are comparable to that of the MW, with 
log M*/M$_{\odot}$=10.7 \citep{flynn2006}.  The X-shaped inner bar
component in MW \citep{nataf2010,ness2012,wegg2013} is also very similar to the X-shaped structures of our sample
galaxies. It has been suggested that MW has no classical bulge.

In our classification the inner component of the bar is either a
barlens or an X-shape. In particular in the overlapping region in Figure 2a,
the assigned class might depend on the strength of the X-shape. Figure 3
shows examples of galaxies with/without prominent X-structures, both
at high and intermediate inclinations.  The upper row shows two
strongly X-shaped bars with $i\approx$ 90$^\circ$ (NGC 3628) and with
$i$=45$^\circ$ (IC 5240). In both cases the X-shape is clear even in
the original image.  The lower row shows a boxy bulge with only a weak
X-shape detected in unsharp mask (NGC 4565, $i\approx$ 90$^{\circ}$) and
a barlens (NGC 4643, $i$=37$^{\circ}$). The barlens has the same appearance
as the boxy bulge (with a weak X-shape) might have at the same
inclination.  Clearly, the classification as barlens or X-shape depends
on both inclination and the prominence of the structure.

 It is of interest to compare our morphological B/P/X/bl frequency
  to earlier studies made using either low or high-inclination
  samples. Erwin $\&$ Debattista (2013) made isophotal analysis for
  moderately inclined galaxies ($i<65^\circ$) in the Hubble type range
  $-3 \leq T \leq 2$ and found that 31$\%$ of barred galaxies show
  boxy or spurs morphology.  Using the same type/inclination range our
  magnitude-limited subsample has an even higher frequency: 46$\%$ of
  barred galaxies (N$_{bar}$=134 out of N$_{tot}$=223) have either
  barlenses or X-shapes (N$_{bl}$=53, N$_X$=9). Relaxing the
  inclination limit gives, as it should, practically similar high
  frequency (51$\%$; N$_X$=35, N$_{bl}$= 55, N$_{bar}$=175).
  Our X/bl frequency is similar to the overall fraction of
  45$\%$ of B/P bulges found by L\"utticke et al. (2000) among the
  edge-on galaxies of all Hubble types.  However, in contrast to
  L\"utticke et al. (2000), our fraction drops to zero beyond
  $T=5$ (10$\%$ for T=3-5), wherease in their study the fraction stays
  over 40$\%$ even for $T=7$, probably due to weak B/Ps becoming dominant.

\subsection{Decomposing the relative fluxes of bulge/bl/B/P/X-shaped structures}

If bars (thin bar + bl/X-structure) can 
account for
most of the inner surface brightness, we may well ask how much room
is left for massive classical bulges in these systems? We tackle this
question by detailed structural decompositions.

Examples of our decompositions for barlens (NGC 4643) and X-shaped (IC
5240) galaxies are shown in Fig. 4. For NGC 4643 a separate bulge
is fitted ($n$=0.7), resulting to $B/T$$\sim$0.1. In IC 5240 the few
central pixels, fitted with the PSF, contain only $\sim$1$\%$ of the
flux, indicating that this galaxy has no separate bulge.  Collecting
all decompositions we find $<B/T>$$\sim$0.1 and $n$$\sim$1.5 for both
types. As a large majority of the decomposed galaxies are classified as
early-type disks this is a very small number, corresponding to that
typical for Sbc-Sc spirals (Laurikainen et al. 2010). For
  comparison, if the bl/X-shaped component were omitted from an
  otherwise similar decomposition, this results in
  $<B/T>$=0.35, similar to $B/T$'s previously obtained
  by Gadotti (2009) and Laurikainen et al. (2007,
  2010)\footnote{It has been shown previously that completely
  omitting the bar in the fit would further increase the relative mass
  erroneously assigned to the bulge (Laurikainen et al. 2006).},
though Laurikainen et al. (2007, 2010) made also more complex decompositions
resulting to lower $B/T$-values.

In the current decompositions
barlenses and X-shaped components account for
  relative fluxes of $<$F(bl)/F(tot)$>$ = 0.18$\pm$0.11 and
  $<$F(X)/F(tot)$>$ = 0.08$\pm$0.02. 
The fact that barlenses appear  marginally more prominent
could be a selection effect: even weak X-shapes stand out in edge-on
galaxies, whereas a small barlens is hard to identify reliably and is thus
more easily discarded from classification.


In Figure 5 we further show that both F(bl)/F(thin bar) and
F(bl)/F(tot) correlate with bar strength (A2). Also, barlenses seem to
be fairly round, e.g. they have large axial ratios.  Clearly, a
prominent barlens means also a strong bar.  These correlations,
  as well as the rather round appearance of barlenses, are predicted
  by those simulation models in Athanassoula et al. (2013,
  2014), which start from initial conditions including gas.

\section{Summary and implications}

We have shown observational evidence that barlenses are the more
  face-on counterparts of the vertically thick B/P/X-shaped bar
  structures, in a good agreement with the simulations by Athanassoula
et al. (2014).  Consistent with this idea is that the host galaxy
  properties of both structures are similar, having similar masses and
  red B-3.6$\mu$m colors. Also, together the two types of galaxies
  form a flat axial ratio distribution.  Barlenses appear
  shifted towards earlier Hubble types, most likely because in galaxy
  classifications they are mistaken with classical bulges.
  Accounting for their central components as bulges leads
  to $<B/T>$$\sim$0.1 and $<n>$$\sim$1.5 in both types, inconsistent
  with the classical bulge picture. 
According to our analysis, $\sim 50\%$ of early type ($-3 \leq T \leq 2$) bars 
 contain a B/P/X/bl structure.


Our results are consistent with many observations of bulges, which are
often interpreted to cover the same galaxy regions as the vertically
thick inner bar components. Almost all bulges are fast rotating
(Cappellari et al. 2011, 2013). Many explanations exist for that,
but the observation fits also to our interpretation where the vertically thick inner bar regions
form part of the bar, rotating in a similar manner as the 
rest of the disk.  The observation that bulges in
the early-type disk galaxies are generally red \citep{driver2006} also
naturally fits into our interpretation, because bars often have old
stellar populations \citep{perez2009}. The major implications are:

1. {\it Milky Way mass galaxies having no classical bulges are common
  in the nearby Universe.}  This result seems to hold even for
early-type disk galaxies where the most massive classical bulges are
assumed to reside.

2. {\it Our results put strong constrains to those galaxy formation
  models in which massive classical bulges form in the early
  Universe}. Either the massive classical bulges formed at high
redshifts have been destroyed at some stage of galaxy evolution, or in
many barred galaxies they never formed.

\section*{Acknowledgments}

EL and HS acknowledge financial support from the Academy of
Finland, EA and AB from the CNES, and all from EU grant
PITN -GA-2011-289313 (DAGAL EU-ITN network). We thank the
anonymous referee of valuable comments.

\begin{figure}
{\includegraphics[angle=0,width=7.4cm]{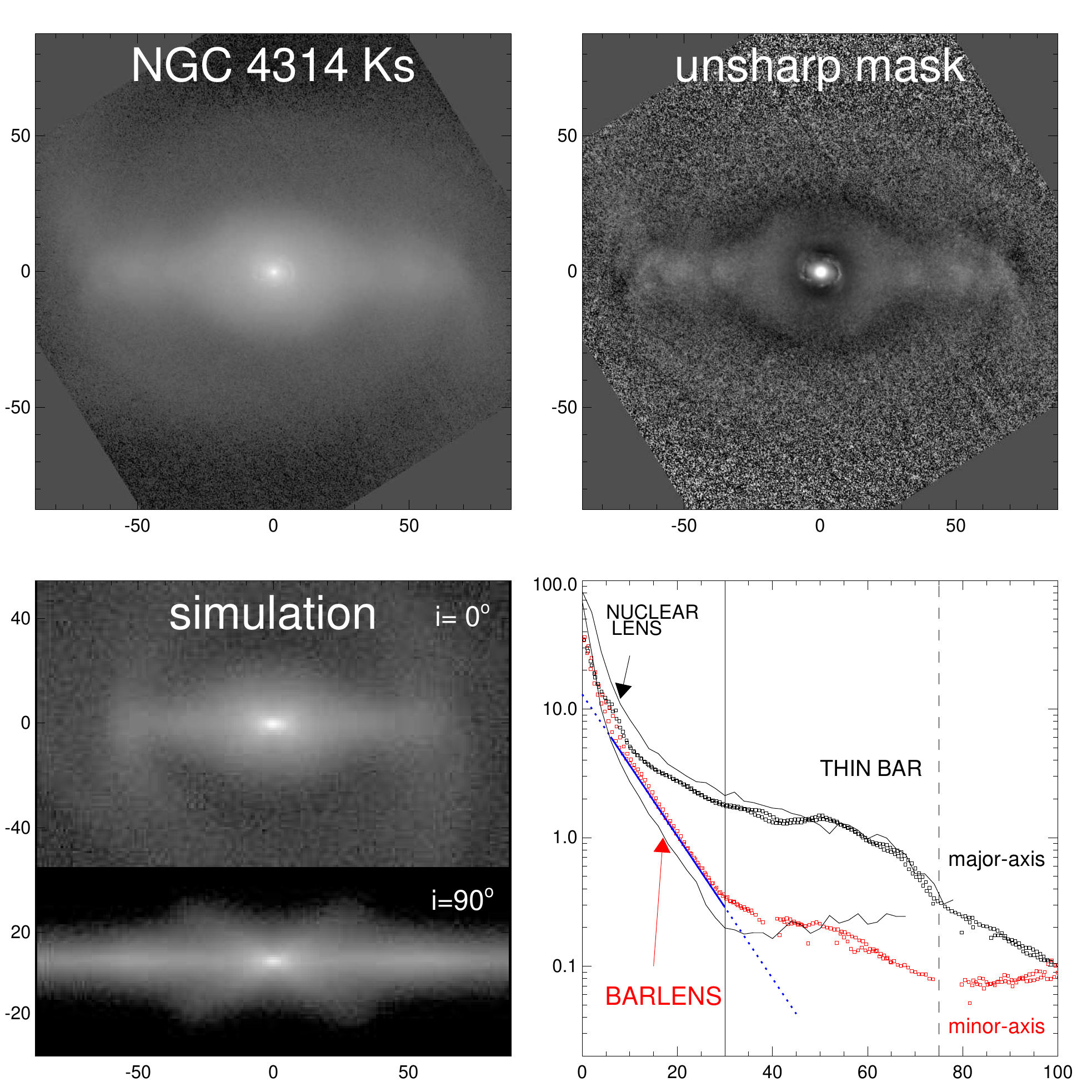}}

 \caption{An example of a barlens galaxy NGC 4314, showing the
   original K$_s$-band image (from Laurikainen et al. 2011), the
   unsharp mask image, and the surface brightness profile along the
   bar major and minor axis (symbols)
The lower left panel shows the simulation model gtr115
   from Athanassoula et al. (2013), both in face-on and edge-on
   view. The simulation model profiles are also shown by solid lines in the profile plot.
  Axis labels are in arcseconds in all panels. }
\end{figure}

\begin{figure}
\begin{centering}
\includegraphics[angle=0,width=6.5cm]{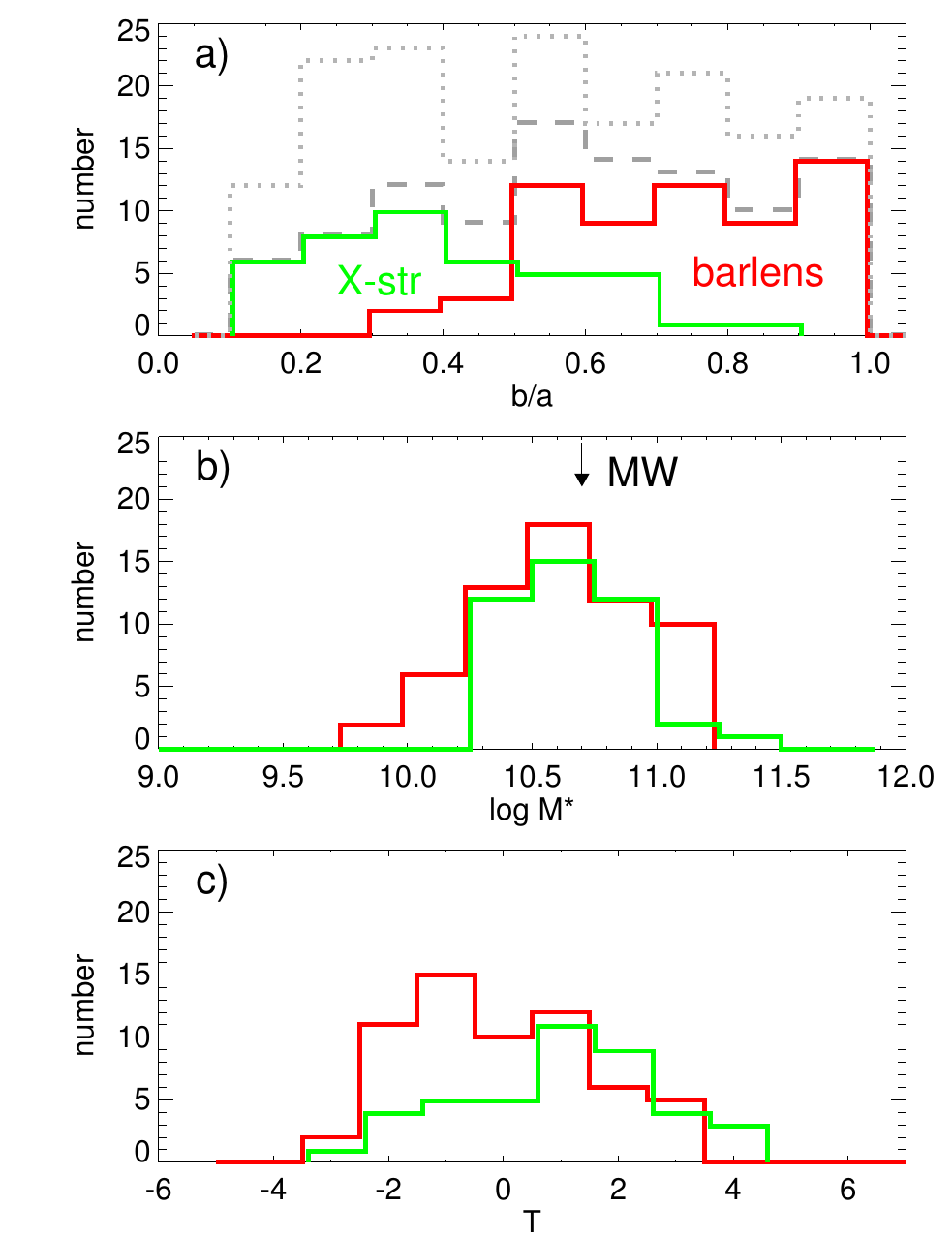}

 \caption{The distributions of (a) galaxy minor-to-major (b/a) axis
   ratios, (b) total stellar masses (in units of solar masses) and (c)
   Hubble types T of the galaxies hosting  either barlens (red) or
   X-shaped structures (green). In these plots the magnitude-limited (B$_T \leq$12.5 mag) subsample is used ($N_{tot}=597,N_{bar}=365,N_X={42},N_{bl}=61$). In a) also the combined barlens+X-shape distribution is shown, both for the above sample
 (dashed line) and for the original sample of N=2465 galaxies. The dotted line shows all bl+X-shapes
of the full combined sample.
%
%
The stellar masses are calculated as in
Munoz-Mateos et al.  (2013) for S$^4$G. Note that the potentially
  missing gas-poor, early type, $i>65^\circ$   galaxies with 
  X-shapes would further flatten the b/a distribution.
}
\end{centering}
\end{figure}


\begin{figure}
\begin{centering}
\hskip -0.3cm
\includegraphics[angle=0,width=9cm]{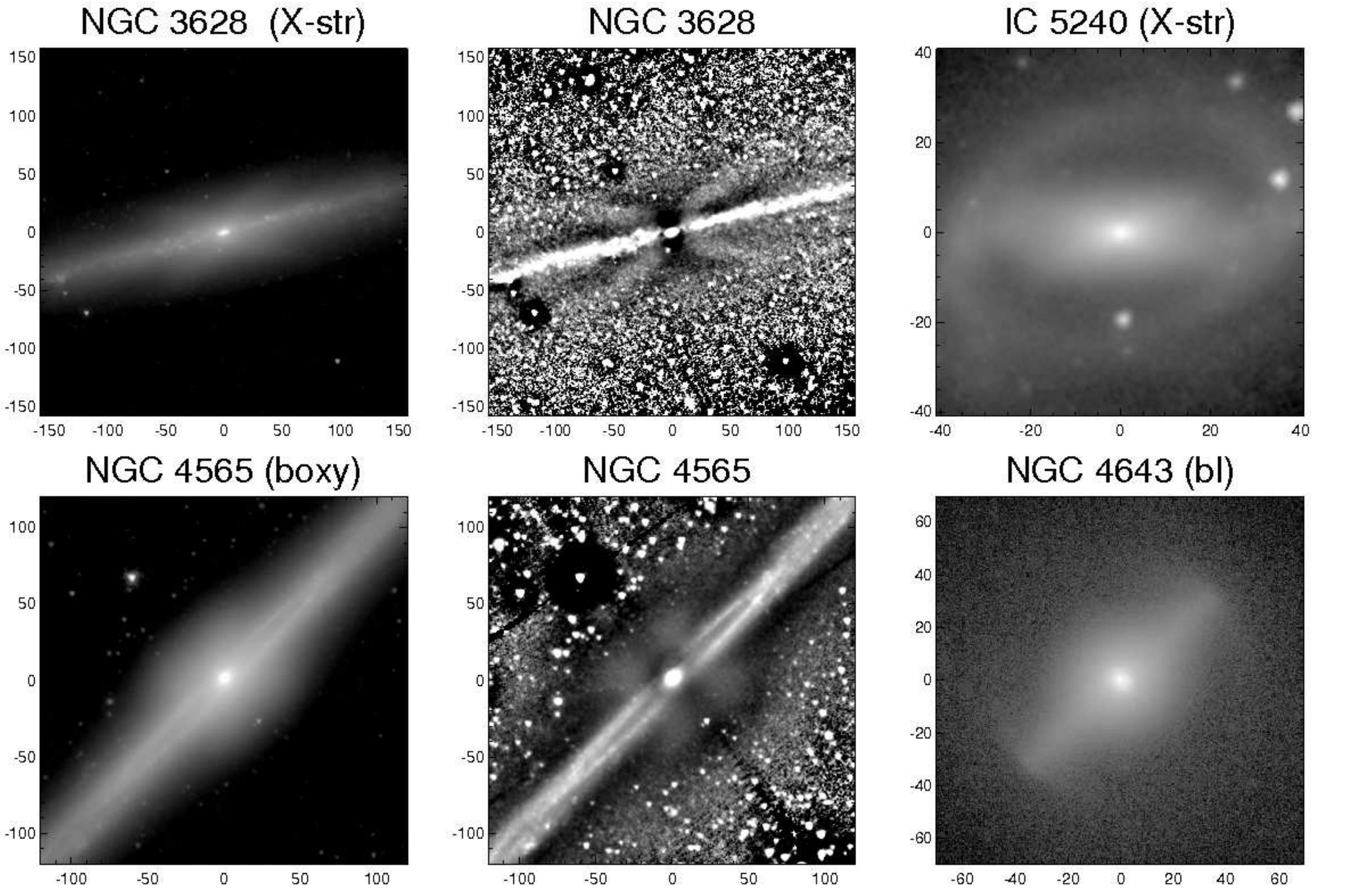}

 \caption{Upper row shows two X-shaped galaxies, NGC 3628
   and IC 5240, and the lower row boxy/barlens galaxies, NGC 4565 and
   NGC 4643.  In the left panels the galaxies are nearly edge-on,
   whereas in the right panels the inclinations are 45$^\circ$
   and 37$^\circ$ for IC 5240 and NGC 4643, respectively. The middle panels
   show the unsharp mask images of the edge-on galaxies. The unsharp masks
for IC 5240 and NGC 4643 are shown in Figure 4. The images of NGC 4643 and
IC 5240 are from NIRS0S, the other two galaxies are from S$^4$G.}
\end{centering}
\end{figure}

\begin{figure}
\begin{centering}
{
\includegraphics[angle=0,width=9.5cm]{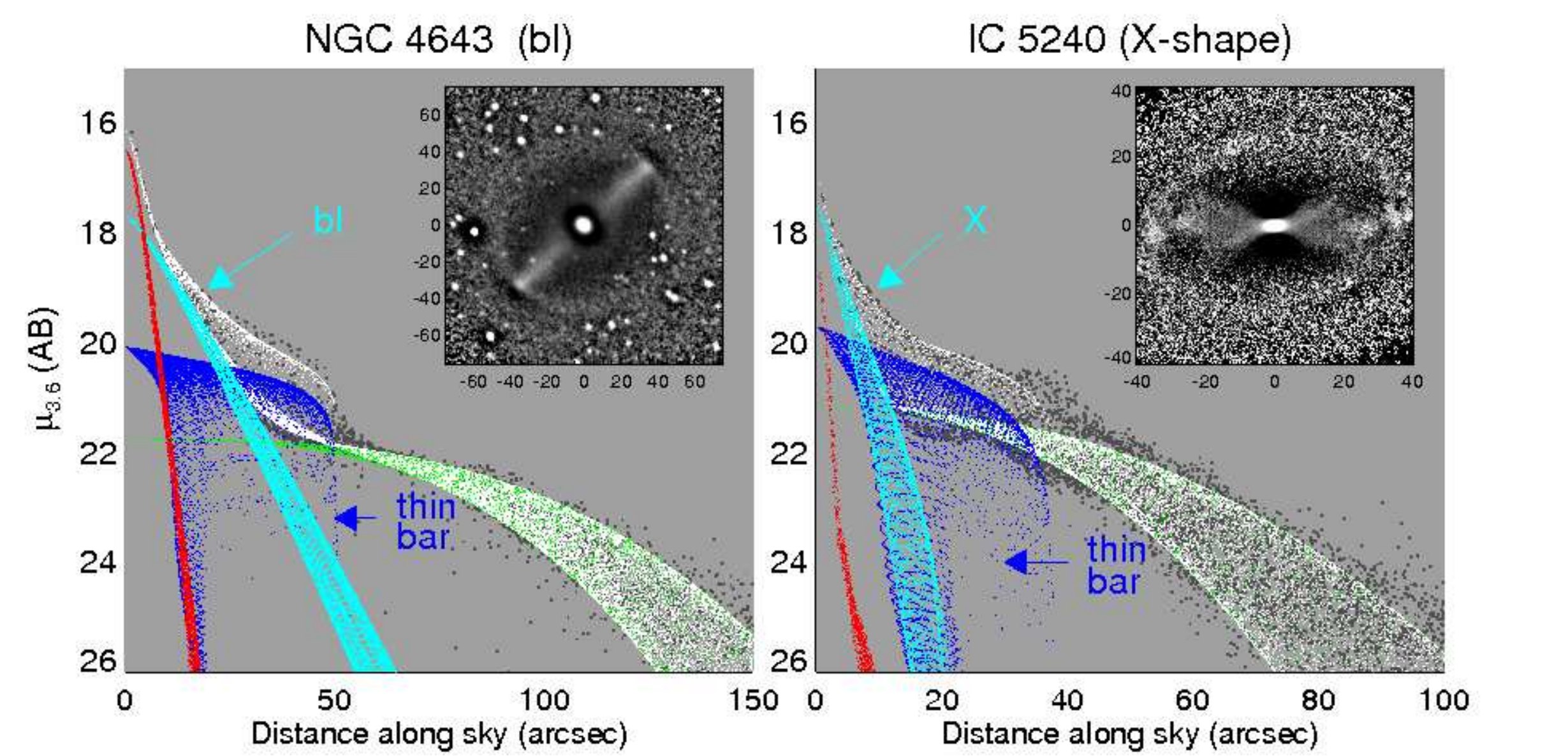}}
 \caption{Decomposition models for the barlens and X-shaped galaxies
   NGC 4643 and IC 5240, for which galaxies the images are shown in Figure
   3. Black dots are the pixel values of the two-dimensional images,
   and white dots show the pixel values of the decomposition
   models. Red and green dots show the bulge and the disk components,
   whereas the dark and light blue indicate the thin and thick bar (bl or X)
   components. See the details of the models in the text. The inserts
   show the unsharp mask images. These decompositions use 3.6$\mu$m images,
but the bulge parameters are similar when using K$_s$-band images. 
The mean Hubble types
for the decomposed bl and X-shaped galaxies are $<$T$>$=-0.75 and 1.2, respectively.
}
\end{centering}
\end{figure}


\begin{figure}
\begin{centering} \includegraphics[angle=0,width=6cm]{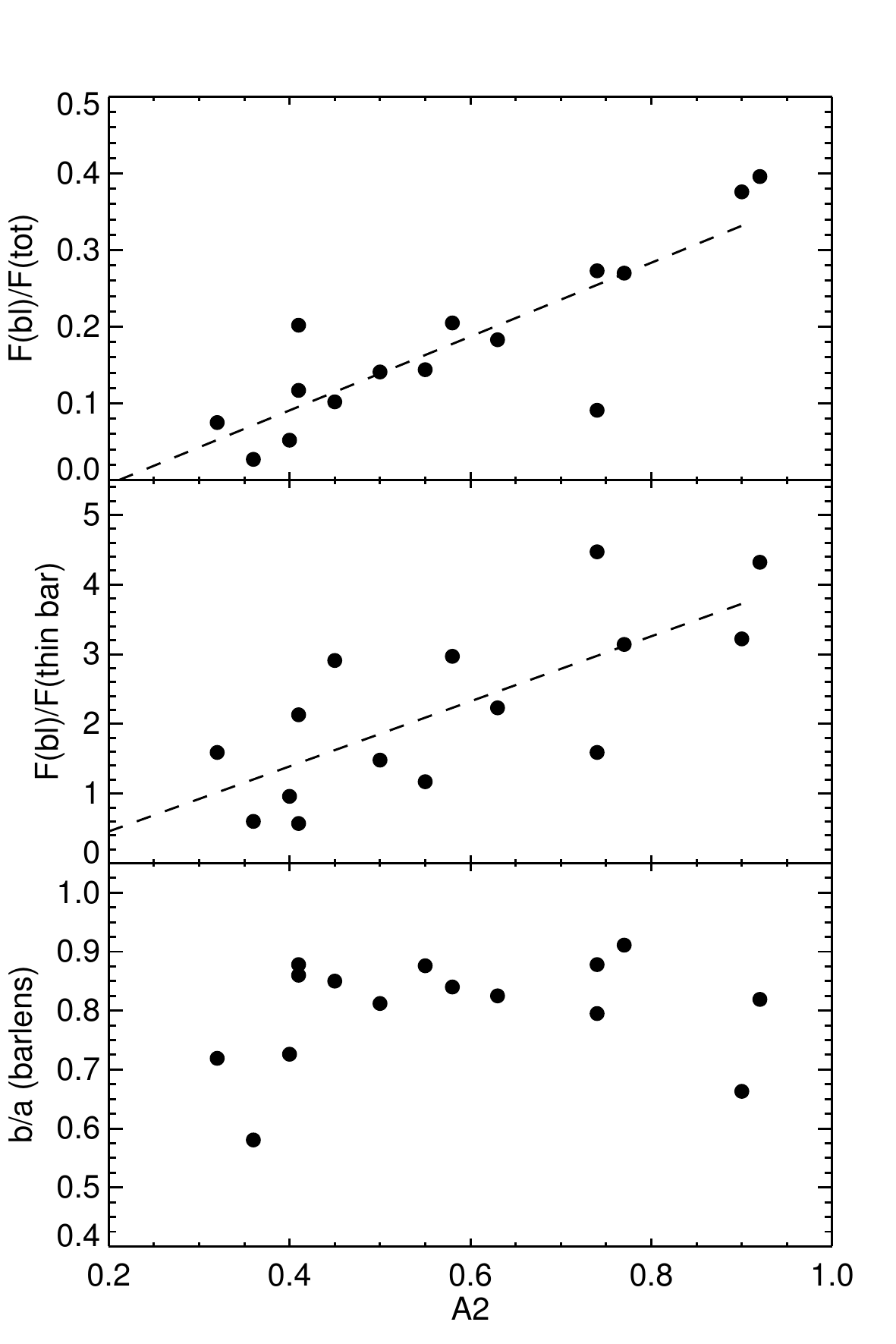}
 \caption{Observed relative fluxes of barlenses are shown as a
   function of A2, for all those barlens galaxies for which decompositions 
   were made in this study. A2 is the peak value of the m=2 Fourier
   amplitude of density in the bar region. In the upper panel the flux is
   normalized to the total galaxy flux (F(bl)/F(tot)), and in the
   middle panel to the flux of the thin bar component (F(bl)/F(thin
   bar)). Lower panel shows the observed axial ratio of the barlens,
   b/a, as a function of A2 for the same galaxies. The axial ratios
   are from Laurikainen et al. (2011). }
\end{centering}
\end{figure}

\bsp

\label{lastpage}

\end{document}